\documentclass[runningheads]{llncs}
\usepackage[T1]{fontenc}
\usepackage{graphicx}
\usepackage{tikz-cd}
\tikzcdset{diagrams={column sep=8ex, row sep=5ex}} 
\usepackage{tikz}
\usetikzlibrary{babel,automata, positioning, arrows}

\usepackage{amssymb}
\usepackage{amsmath}
\usepackage{amsthm}
\usepackage{bussproofs}
\usepackage{braket}

\usepackage{enumerate}
\usepackage{enumitem}

\usepackage[british]{babel}
\usepackage{hyperref}
\usepackage[capitalise]{cleveref}

\usepackage{float}
\usepackage{tcolorbox}
\usepackage{todonotes}

\usepackage{color}

\newcommand{\Id}{\mathsf{Id}}
\newcommand{\id}{\mathsf{id}}

\newcommand{\D}{\mathcal{D}}

\newcommand{\pow}{\mathcal{P}}

\newcommand\CSet{\mathsf{Set}}

\newcommand\Rel{\mathsf{Rel}}

\newcommand\supp{\mathrm{supp}}
\newcommand{\reca}[2]{\not\asymp_{#1}^{#2}}

\newcommand\eqdef\triangleq
\newcommand{\apart}[1]{\mathrel{\#_{#1}}}
\newcommand\subobj\rightarrowtail

\newcommand\fin\overline

\newcommand\paren[1]{\left({#1}\right)}
\newcommand\tuple\paren%

\begin{document}
\title{Proving Behavioural Apartness\thanks{%
This research is partially supported by the NWO grant No.~OCENW.M20.053, 
the EPSRC NIA Grant EP/X019373/1, and the Royal Society Travel Grant 
IES\textbackslash R3\textbackslash223092.}}
%
%
\author{Ruben Turkenburg\inst{1}\orcidID{0000-0001-7336-9405} \and
	Harsh Beohar\inst{2}\orcidID{0000-0001-5256-1334} \and
	Clemens Kupke\inst{3}\orcidID{0000-0002-0502-391X} \and
	Jurriaan Rot\inst{1}\orcidID{0000-0002-1404-6232}}
\authorrunning{R. Turkenburg et al.}
%
\institute{Radboud University, Nijmegen, Netherlands \\
	\email{\{ruben.turkenburg@,jrot@cs.\}ru.nl} \and
	University of Sheffield, United Kingdom \\
	\email{h.beohar@sheffield.ac.uk} \and
	Strathclyde University, United Kingdom \\
	\email{clemens.kupke@strath.ac.uk}}
\selectlanguage{british}
\frenchspacing
\maketitle              
\begin{abstract}
	Bisimilarity is a central notion for coalgebras. In recent work, Geuvers and Jacobs suggest to focus on apartness, which they define by dualising coalgebraic bisimulations. This yields the possibility of finite proofs of distinguishability for a wide variety of state-based systems.

	We propose \emph{behavioural apartness}, defined by dualising behavioural equivalence rather than bisimulations. A motivating example is the subdistribution functor, where the proof system based on bisimilarity requires an infinite quantification over couplings, whereas behavioural apartness instantiates to a finite rule.
	In addition, we provide optimised proof rules for behavioural apartness and show their use in several examples.
\end{abstract}
\section{Introduction}
In the study of coalgebra as a general approach to state-based systems, a number of notions of equivalence of states have been developed. Among these are: bisimilarity as defined by Aczel \& Mendler; behavioural equivalence; and bisimilarity defined using (canonical) relation lifting due to Hermida \& Jacobs~\cite{DBLP:conf/ctcs/AczelM89,DBLP:journals/iandc/HermidaJ98}.
Alongside these definitions, proof systems have been developed for some of these notions, to give a syntactic way to deduce equivalence of states in a coalgebra. However, due to the coinductive nature of coalgebraic notions of equivalence, these are challenging to develop as they have to deal with the circularity inherent in coinduction~\cite{DBLP:conf/calco/LucanuGCR09,DBLP:conf/fossacs/CloustonBGB15,basold2018mixed,DBLP:conf/calco/SprungerM17}.

Another (closely related) line of research is coalgebraic modal logic~\cite{DBLP:journals/apal/Moss99}, where coalgebras give the semantics of modal formulas. These logics bring with them another notion of equivalence, namely we can call states logically equivalent if they satisfy exactly the same formulas of a given logic. Comparing these equivalences to the aforementioned coinductive ones, we can ask whether bisimilar states are logically equivalent (adequacy) and whether logically equivalent states are bisimilar (expressivity). Already (implicitly) present in the study of these properties are dual notions, defining states instead as inequivalent (see, e.g.,~\cite{DBLP:journals/iandc/DesharnaisEP02,DBLP:conf/icalp/FijalkowKP17,DBLP:conf/cmcs/0001MS20,DBLP:journals/lmcs/WissmannMS22}). For example, expressivity of a logic may more easily be shown by giving, for any pair of non-bisimilar states, a formula distinguishing them.

In recent work, Geuvers \& Jacobs~\cite{DBLP:journals/lmcs/GeuversJ21} focus on such notions of inequivalence (or \emph{distinguishability}), and investigate what they call apartness. This is done first for concrete examples, such as labelled transition systems, before they show how their definitions can be obtained coalgebraically. For this, the starting point is bisimilarity defined using (canonical) relation lifting, which for a coalgebra (in the category \(\CSet\)) is the largest relation \(R\) such that if two states are related, then their successor structures must be related by the relation lifting of the behaviour functor applied to \(R\). The dual notion of apartness, again on a \(\CSet\) coalgebra, can then be described as the smallest relation \(R\) such that if successor structures of two states are related by the relation obtained by applying \emph{opposite} relation lifting to \(R\), then \(R\) must relate those two states. This immediately gives us that states are bisimilar if and only if they are not apart.
Furthermore, in \emph{op.~cit.}, it is shown how their notion of apartness quite directly gives rise to a derivation system allowing apartness to be determined inductively.

In this work, we continue the investigation of the possible notions of inequivalence related to existing definitions of bisimilarity and behavioural equivalence, and the development of corresponding proof systems. We start by recalling the apartness notion due to Geuvers \& Jacobs, before giving an example of a functor for which this does not easily provide a satisfactory proof system for apartness (\cref{sec:cobisimilarity}). Namely, the subdistribution functor, which gives a rule whose premise universally quantifies over the infinite set of couplings. This does not allow apartness to be proved by giving some witness, which can be done for the systems considered in~\cite{DBLP:journals/lmcs/GeuversJ21}.

An alternative characterisation of bisimilarity is present in work on probabilistic systems based on summing successor distributions over equivalence classes~\cite{DBLP:journals/iandc/LarsenS91}, which has further been shown to coincide with coalgebraic notions of bisimilarity and behavioural equivalence~\cite{DBLP:journals/tcs/VinkR99,sokolova2005coalgebraic,DBLP:journals/tcs/Sokolova11}.
We use this characterisation from the starting point of behavioural equivalence to give a dual notion, which we call \emph{behavioural apartness}. This gives rise to a proof system where the inductive step does not involve a universal quantification, allowing finite proofs of apartness. We present this system, prove it to be sound and complete with respect to the dual of behavioural equivalence for finitary functors on \(\CSet\), and provide optimised rules which we compare to the basic rule by application to examples (\cref{sec:behap,sec:optimisations}).

\subsubsection{Notation}
For an equivalence relation $R \subseteq X \times X$, we denote the quotient map by \(q_R \colon X \to X/R\), which sends an element \(x\) to its equivalence class \({[x]}_{R}\). We extend this to arbitrary relations $R \subseteq X \times X$, by defining $q_R$ as $q_{e(R)}$, where $e(R)$ is the equivalence closure of $R$.
Further, we write \(\overline{R}\) for the complement of a relation \(R \subseteq X \times X\) on a set \(X\).

\section{Cobisimilarity}\label{sec:cobisimilarity}
We start by recalling the well-known notion of coalgebraic bisimulation, defined via relation lifting~\cite{DBLP:journals/iandc/HermidaJ98}. The dual of this notion has been investigated in~\cite{DBLP:journals/lmcs/GeuversJ21}, using the so called fibred opposite construction. In this section, we review this approach and show how it can be utilised to obtain a proof system for apartness. We will further show that there exist functors for which this dualised definition does not give rise to a satisfactory proof rule for cobisimilarity, in contrast to the rules provided in \emph{op.~cit.}

For an endofunctor \(B\) on the category \(\CSet\) of sets and functions, the (canonical) relation lifting \(\Rel(B)\) is defined on a relation \(R \subseteq X \times X\) as follows
\[
	\Rel(B)(R) := \{ (s,t) \mid \exists z \in B(R) \ldotp B (\pi_1) (z) = s \land B (\pi_2) (z) = t \}.
\]
This is then used to define the following notion of coalgebraic bisimulation:
\begin{definition}
A relation \(R \subseteq X \times X\) is a coalgebraic bisimulation for a coalgebra \(\gamma \colon X \to B(X)\) if $R \subseteq {(\gamma \times \gamma)}^{-1} \circ \Rel(B)(R)$, that is, it satisfies:
\begin{equation*}
	\AxiomC{\(x \mathrel{R} y\)}
	\UnaryInfC{\(\gamma(x) \mathrel{\Rel(B)(R)} \gamma(y)\)}
	\DisplayProof
\end{equation*}
The largest such relation is called \emph{bisimilarity}, denoted by \(\underline{\leftrightarrow}\).
\end{definition}
Replacing \(R\) above with its complement and taking the contrapositive, we obtain a dual definition.
\begin{definition}
A \emph{cobisimulation} on \(\gamma \colon X \to B(X)\) is a relation \(R\) satisfying
\begin{equation}\label{eq:cobisim}
	\AxiomC{\(\gamma(x) \mathrel{\overline{\Rel(B)(\overline{R})}} \gamma(y)\)}
	\UnaryInfC{\(x \mathrel{R} y\)}
	\DisplayProof
\end{equation}
	Equivalently, \(R\) is a cobisimulation if $R \supseteq {(\gamma \times \gamma)}^{-1} \circ \overline{\Rel(B)(\overline{R})}$, where the right-hand relation is constructed as the following composition of monotone maps:
\begin{equation*}
\begin{tikzcd}
	\Rel_X \arrow[r,"\overline{(-)}"] & \Rel_X \arrow[r,"\Rel(B)"] & \Rel_{B(X)} \arrow[r,"\overline{(-)}"] & \Rel_{B(X)} \arrow[r,"(\gamma \times \gamma)^{-1}"] & \Rel_X
\end{tikzcd}
\end{equation*}
where we use \(\Rel_X\) to denote the lattice of relations on \(X\), ordered by inclusion.
\end{definition}
The relation of \emph{cobisimilarity} is then defined as the smallest cobisimulation. We denote this relation using \(\underline{\apart{}}\), and we see that \(\underline{\leftrightarrow} = \overline{(\underline{\apart{}})}\).
For a more general coalgebraic treatment, see~\cite{DBLP:journals/lmcs/GeuversJ21}. Note that we differ in terminology from \emph{op.~cit.} in using \emph{cobisimilarity} rather than apartness, to distinguish from both the dual of equivalence relations, and the notion of behavioural apartness introduced later.

\begin{example}[Labelled Transition Systems]
In~\cite{DBLP:journals/lmcs/GeuversJ21} cobisimilarity dual to weak forms of bisimulation are studied. To simplify the presentation, we instantiate the definition of cobisimulation to LTSs, which we model as coalgebras for the functor \(\pow{(-)}^A\) for an input alphabet \(A\).

The relation lifting of \(\pow{(-)}^A\) acts as follows
\begin{align*}
	\Rel(\pow{(-)}^A)(R) = \{ (f,g) \mid \forall a \in A \ldotp & [\forall x \in f(a) \ldotp \exists y \in g(a) \ldotp (x,y) \in R]    \\
	\land                                                       & [\forall y \in g(a) \ldotp \exists x \in f(a) \ldotp (x,y) \in R] \}
\end{align*}
Applying the dualisation as in~\eqref{eq:cobisim}, we see that a relation \(R\) is a cobisimulation on a coalgebra  \(\gamma \colon X \to \pow{(X)}^A\) if it satisfies the following two rules
\begin{align*} 
	\AxiomC{\(x \stackrel{a}{\to} x'\)}
	\AxiomC{\(\forall y' \ldotp y \stackrel{a}{\to} y' \implies x' \mathrel{R} y'\)}
	\BinaryInfC{\(x \mathrel{R} y\)}
	\DisplayProof
	\quad
	\AxiomC{\(y \stackrel{a}{\to} y'\)}
	\AxiomC{\(\forall x' \ldotp x \stackrel{a}{\to} x' \implies x' \mathrel{R} y'\)}
	\BinaryInfC{\(x \mathrel{R} y\)}
	\DisplayProof
\end{align*}
where \(x \stackrel{a}{\longrightarrow} x'\) means \(x' \in \gamma(x)(a)\).
Then cobisimilarity on LTSs is the smallest relation satisfying these rules. This gives an inductive proof system for cobisimilarity, as further explained in~\cite{DBLP:journals/lmcs/GeuversJ21}.

We finish this example by giving a derivation of cobisimilarity for the states \(x\) and \(y\) in the following LTS:
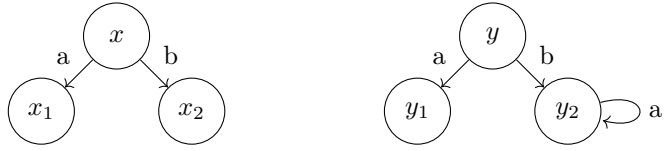
\begin{center}
	\begin{tikzpicture}[auto,on grid]
		\node[state] (x) {\(x\)};
		\node[state] (x1) [below left=of x] {\(x_1\)};
		\node[state] (x2) [below right=of x] {\(x_2\)};
		\node[state] (y) [right=5cm of x] {\(y\)};
		\node[state] (y1) [below left=of y] {\(y_1\)};
		\node[state] (y2) [below right=of y] {\(y_2\)};

		\path[->]
		(x) edge node[swap] {a} (x1)
		edge node {b} (x2)
		(y) edge node[swap] {a} (y1)
		edge node {b} (y2)
		(y2) edge[loop right] node {a} (y2)
		;
	\end{tikzpicture}
\end{center}
This derivation goes as follows:
\begin{align*}
	\AxiomC{\(x \stackrel{b}{\to} x_2\)}
	\AxiomC{\(y_2 \stackrel{a}{\to} y_2\)}
	\AxiomC{}
	\UnaryInfC{\(\forall x' \ldotp x_2 \stackrel{a}{\to} x' \ldotp y_2 \apart{} x'\)}
	\BinaryInfC{\(x_2 \apart{} y_2\)}
	\UnaryInfC{\(\forall y' \ldotp y \stackrel{b}{\to} y' \ldotp x_2 \apart{} y'\)}
	\BinaryInfC{\(x \apart{} y\)}
	\DisplayProof
\end{align*}
The states \(x\) and \(y\) can be distinguished by their b-transitions, and their successors by the presence of an a-transition only on \(y_2\).
\end{example}

\begin{example}[Subdistributions]\label{sec:relsubdists}
We continue by instantiating the rule~\eqref{eq:cobisim} for cobisimulations to coalgebras for the subdistribution functor. In contrast to LTSs above, the relation lifting approach does not give a pleasant proof system.

The (finitely supported) subdistribution functor \(\D_s \colon \CSet \to \CSet\) is defined as follows, where \(\supp(\mu) = \{ x \in X \mid \mu(x) \neq 0 \}\).
\[
	\D_s (X) = \left\{ \mu \colon X \to [0,1] \mid \sum_{x \in X} \mu(x) \leq 1, \supp(\mu) \text{ finite} \right\}
\]
We may equivalently write such distributions as formal sums \( \sum_{x \in X} \mu(x) \ket{x} \).
This allows us to denote the functor's action on arrows as:
\[
	\D_s(f \colon X \to Y) \colon \left(\sum_{x \in X} \mu(x) \ket{x}\right) \mapsto \left(\sum_{x \in X} \mu(x) \ket{f(x)}\right)
\]

We can now elaborate the action of the relation lifting of the subdistribution functor \(\D_s\) on a relation \(R \subseteq X \times X\):
\begin{align*}
	\Rel(\D_s)(R) & = \{ (\D_s (\pi_1) (z), \D_s (\pi_2) (z)) \mid z \in \D_s (R) \}                                                                                                                          \\
	              & = \left\{ \left[ \sum_{x \in X} \left[ \sum_{y \in X} \mu(x,y) \right] \ket{x} , \sum_{y \in X} \left[ \sum_{x \in X} \mu(x,y) \right] \ket{y} \right] \mid \mu \in \D_s (R) \right\}
\end{align*}
Here, \(\mu\) is a subdistribution over the elements of \(R\), and the corresponding pair in \(\Rel(\D_s)(R)\) consists of the left and right marginals of \(\mu\).

Now, we can instantiate the premise of the rule for cobisimulations for this relation lifting on a coalgebra \(\gamma \colon X \to \D_s(X)\) to obtain the following rule:
\begin{equation}\label{eq:couplings}
	\AxiomC{\(\forall \mu \in \D_s (\overline{R}) \ldotp \gamma(x) \neq \D_s \pi_1 (\mu) \lor \gamma(y) \neq \D_s \pi_2 (\mu)\)}
	\UnaryInfC{\(x \mathrel{R} y\)}
	\DisplayProof
\end{equation}
However, this quantifies over the infinite set of subdistributions on the complement of \(R\). This is rather unfortunate since we would like to prove apartness by giving a ``witness'', as was the case for LTSs. There, apartness of states \(x,y\) can be shown by giving a successor of \(x\) which is apart from all successors of \(y\).

	If we consider the example in~\eqref{fig:subMC}, we see that \(x\) and \(y\) are cobisimilar, as they have different transition weights to sets of equivalent states. To see this using the rule~\eqref{eq:couplings}, we could try to reason that given a distribution \(\mu \in \D_s(\overline{R})\), the weight of the pair \((x_1, y_1)\) (two states which are clearly equivalent) will always be such that the left and right marginals do not match the transition probabilities from \(x\) to \(x_1\) and from \(y\) to \(y_1\), as \(\mu\) should not assign mass to the pair \((x_1,y_2)\) of inequivalent states.
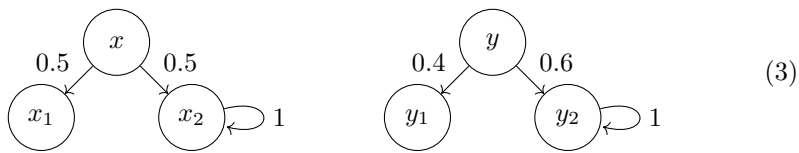
\begin{equation}\label{fig:subMC}
	\begin{tikzpicture}[auto,on grid,baseline=(current bounding box.center)]
		\node[state] (x) {\(x\)};
		\node[state] (x1) [below left=of x] {\(x_1\)};
		\node[state] (x2) [below right=of x] {\(x_2\)};
		\node[state] (y) [right=5cm of x] {\(y\)};
		\node[state] (y1) [below left=of y] {\(y_1\)};
		\node[state] (y2) [below right=of y] {\(y_2\)};
		\path[->]
		(x) edge node[swap] {\(0.5\)} (x1)
		edge node {\(0.5\)} (x2)
		(y) edge node[swap] {\(0.4\)} (y1)
		edge node {\(0.6\)} (y2)
		(x2) edge[loop right] node {\(1\)} (x2)
		(y2) edge[loop right] node {\(1\)} (y2)
		;
	\end{tikzpicture}
\end{equation}
\end{example}

We see two main issues with rule~\eqref{eq:couplings}: the reasoning provided (where we choose the pair \((x_1,y_1)\)) is not well reflected in the proof system; and the rule requires reasoning about an infinite set even for a simple \emph{finite} coalgebra. This motivates much of the coming work, in which we move to an apartness notion defined dually to behavioural equivalence rather than coalgebraic bisimulation. The corresponding proof system will both exhibit the desired existential reasoning, and allow an optimisation giving finite proofs of apartness for both finite and infinite systems.

In the next section, we show the dualisation of behavioural equivalence, before instantiating to examples which will already illustrate the benefit of this approach over cobisimilarity.
In \cref{sec:optimisations}, we go on to present the optimised proof rule, and show how this is beneficial by application to the examples of labelled Markov processes and stream systems.

\section{Behavioural Apartness}\label{sec:behap}
In this section, we use a characterisation of behavioural equivalence based on so called \emph{precongruences} due to Aczel \& Mendler~\cite{DBLP:conf/ctcs/AczelM89} to present a basic proof system for the dual: \emph{behavioural apartness}. We will further apply the proof system to coalgebras for the subdistribution functor to show its benefits over the system provided in~\cref{sec:cobisimilarity}.
Throughout this paper we work with coalgebras living in the category \(\CSet\) for an endofunctor $B\colon \CSet\to\CSet$.

\subsection{Behavioural Equivalence}
We start with the general definition of behavioural equivalence on \(\CSet\) coalgebras, based on cospans of coalgebra homomorphisms.

\begin{definition}
	States \(x,y\) of a coalgebra \(\gamma \colon X \to B(X)\) are said to be behaviourally equivalent if there is a coalgebra \(\delta \colon Y \to B(Y)\) and coalgebra morphisms \(f,g \colon (X,\gamma) \to (Y,\delta)\), such that \(f(x) = g(y)\).
\end{definition}

Now the following definition (from~\cite{DBLP:conf/ctcs/AczelM89}) will give an alternative characterisation of behavioural equivalence which is amenable to dualisation:
\begin{definition}\label{def:precong}
	A precongruence on a coalgebra \(\gamma \colon X \to B(X)\) is a relation \(R \subseteq X \times X\) satisfying the following rule:
\begin{equation}\label{eq:precong}
	\AxiomC{\(x \mathrel{R} y\)}
	\UnaryInfC{\(B q_R(\gamma(x)) = B q_R(\gamma(y))\)}
	\DisplayProof
\end{equation}
\end{definition}
Following Aczel \& Mendler we call a precongruence which is an equivalence relation a \emph{congruence}. They further showed that \(R\) is a congruence if and only if it is the kernel of a coalgebra morphism, so that the maximal (pre)congruence is equal to behavioural equivalence (also see~\cite{Gumm99}). The following lemma tells us that this maximal congruence is a greatest fixed point of the operator (implicit in \cref{def:precong}) mapping a relation \(R\) to the kernel \(\ker(Bq_{R} \circ \gamma)\)~\cite[Lemma 4.1]{DBLP:conf/ctcs/AczelM89}.
\begin{lemma}\label{lem:monotone}
	For equivalence relations \(R,S \subseteq X \times X\), we have \(R \subseteq S \implies \forall x,y \in X \ldotp [Bq_R(\gamma(x)) = Bq_R(\gamma(y)) \implies Bq_S(\gamma(x)) = Bq_S(\gamma(y))]\).
\end{lemma}
We note that Staton has shown similar results for more general categories than \(\CSet\)~\cite{DBLP:journals/corr/abs-1101-4223}.

\subsection{Dualising Behavioural Equivalence}
The rule~\eqref{eq:precong} can now be dualised to obtain a definition of behavioural apartness.
\begin{definition}\label{def:behap}
\emph{Behavioural apartness} is the smallest relation satisfying the following rule:
\begin{equation*}
	\AxiomC{\(B q_{\overline{R}}(\gamma(x)) \neq B q_{\overline{R}}(\gamma(y))\)}
	\UnaryInfC{\(x \mathrel{R} y\)}
	\DisplayProof
\end{equation*}
	where \(\overline{R}\) is the complement of the relation \(R\).
\end{definition}

Dually to the case of precongruences and behavioural equivalence, this smallest relation will be an \emph{apartness relation}.
\begin{proposition}
Behavioural apartness is an apartness relation, i.e., it is irreflexive (\(\forall x \in X \ldotp \lnot(x \mathrel{R} x)\)), symmetric, and cotransitive (\(\forall x_1,x_2,x_3 \in X \ldotp (x_1 \mathrel{R} x_2 \implies x_1 \mathrel{R} x_3 \lor x_2 \mathrel{R} x_3)\)).
\end{proposition}

In turning the rule of \cref{def:behap} into a proof system, we make a number of changes. First, we make the proof obligations explicit. Second, we require that all pairs of the relation \(R\) have been proved apart. Finally, we introduce notation for the inequality \(Bq_{\overline{R}} (\gamma(x)) \neq Bq_{\overline{R}}(\gamma(y))\) while also taking the symmetric closure \(R^s\) of \(R\) in the definition, namely we define
\begin{equation}\label{eq:reca}
	t_1 \reca{R}{B} t_2 := Bq_{\overline{R^s}}(t_1) \neq Bq_{\overline{R^s}}(t_2)
\end{equation}
Note that, in this definition, we are taking the equivalence closure of \(\overline{R^s}\), as in \cref{def:precong}. This is equivalent to first taking the \emph{apartness interior} \((R^s)^\circ\) of \(R^s\), defined as the largest apartness relation contained in \(R^s\), and then taking \(\reca{(R^s)^\circ}{B}\) (where the equivalence closure implicit in the definition of \(q\) does not change the relation). Also note that symmetric closure does not commute with complement, i.e., taking the symmetric closure first really gives a distinct inequality. Our motivation for taking this symmetric closure will become clear when we come to examples, where it saves us the work of proving symmetric pairs apart.

Our proof rule for apartness is formally stated in the following theorem. To ensure that finite proof trees suffice, we assume that the functor is finitary. It is well-known that this suffices for behavioural equivalence to converge at $\omega$ in the final chain. In the proof below, we use the results of Worrell on final sequences of finitary $\CSet$ functors to detail the case of behavioural apartness~\cite{DBLP:journals/tcs/Worrell05}.%
\begin{theorem}\label{thm:soundcomp}
	Let \(B\) be a finitary endofunctor on \(\CSet\), and \(\gamma \colon X \to B(X)\) a coalgebra. Then states \(x,y \in X\) are behaviourally apart if and only if we have a proof tree of finite height built from the following rule:
	\begin{equation}\label{eq:genintrule}
		\AxiomC{\(\forall (x',y') \in R \ldotp x' \apart{} y'\)}
		\AxiomC{\(\gamma(x) \reca{R}{B} \gamma(y)\)}
		\BinaryInfC{\(x \apart{} y\)}
		\DisplayProof
	\end{equation}
\end{theorem}
Note that the given ``rule'' is, strictly speaking, a family of rules indexed by relations \(R \subseteq X \times X\). Thus, a proof tree may contain different instances of the rule, involving different choices of \(R\).
\begin{proof}
	The if direction holds by induction, as then \(R\) is contained in behavioural apartness so that \(\overline{R^s}\) contains behavioural equivalence. Using~\cref{lem:monotone}, we then see that \(\gamma(x) \reca{R}{B} \gamma(y) \implies \gamma(x) \reca{\overline{\underline{\apart{}}}}{B} \gamma(y)\), where we write \(\underline{\apart{}}\) for behavioural apartness here. By~\cref{def:behap}, we thus have \(x \mathrel{\underline{\apart{}}} y\).

	For the other direction, first recall that a coalgebra \(\gamma \colon X \to B(X)\) defines a cone \({(f_i \colon X \to B^i 1)}_{i \in I}\) over the final sequence of \(B\) by (transfinite) induction, with \(f_0 \colon X \to 1\) the unique map, \(f_{i+1} = B f_i \circ \gamma\), and for a limit ordinal \(\alpha\), \(f_\alpha = \lim_{i < \alpha} f_i\).
	Futher recall the notion of \(n\)-step behavioural equivalence, defined as the kernel of the \(n\)-th map of the above cone, i.e., \(\ker(f_n)\). This gives a notion of \(n\)-steps behavioural apartness, namely the complement \(\overline{\ker{(f_n)}}\).

	We will now first show that for any \(n < \omega\) and pair of states \((x,y) \in \overline{\ker{(f_n)}}\), there is a proof tree of height \(n\) with \(x \apart{} y\) as its root.
	The base case is trivial, as \(\overline{\ker(f_0)}\) is empty.
	Now suppose that the property holds for some \(N > 0\), and further that we have some \((x,y) \in \overline{\ker{(f_{N+1})}}\). This means (by definition) that
	\[
		Bf_N(\gamma(x)) \neq Bf_N(\gamma(y))
	\]
	We then need to show that there is some set of pairs \(R \subseteq X \times X\) so that
	\begin{equation}\label{eq:apartprem}
		Bq_{\overline{R^s}}(\gamma(x)) \neq Bq_{\overline{R^s}}(\gamma(y))
	\end{equation}
	and we claim that \(\overline{\ker{(f_N)}}\) is such an \(R\), and that it gives us a proof tree of height \(N\) with \(x \apart{} y\) as root.

	By the induction hypothesis, for any \((x',y') \in \overline{\ker{(f_N)}}\), we have proof trees with \( x' \apart{} y'\) at the root, and height \(N\). Further, the premise (\cref{eq:apartprem}) holds by definition of the kernel. Combining these gives us the required proof tree.

	For the limit at \(\omega\), we have
	\[
		\overline{\ker(f^\omega)} = \overline{\left( \bigcap_{i < \omega} \ker(f^i) \right)} = \bigcup_{i < \omega} \overline{\ker(f^i)}
	\]
	so that for \((x,y) \in \overline{\ker(f^\omega)}\), there is some \(i < \omega\) with \((x,y) \in \overline{\ker(f^i)}\). We obtain the required proof tree by (transfinite) induction.

	It is further shown by Worrell~\cite{DBLP:journals/tcs/Worrell05}, that the final sequence for finitary functors on \(\CSet\) stabilises at \(\omega 2\) and, more importantly here, that the connecting maps \(B_\beta \to B_\lambda\) for \(\omega \leq \lambda \leq \beta \leq \omega 2\) are injective. Thus, for \(\omega \leq \beta \leq \omega 2\), we have \(f^\beta(x) \neq f^\beta(y) \implies f^\omega(x) \neq f^\omega(y)\), again giving the required proof tree.
\end{proof}
We note that this could be recovered using results from~\cite{DBLP:journals/mscs/HasuoKC18}, namely that coinductive predicates for finitary functors can be constructed via a final sequence which stabilises after \(\omega\) steps. Dualising, there should be a correspondence between stages of the initial sequence forming proof trees, and the sequence constructing behavioural apartness as an inductive predicate. However, as we work in \(\CSet\), we are able to use the earlier results due to Worrell.

Before we show how rule~\eqref{eq:genintrule} can be improved further, we apply it to the coalgebra for the subdistribution functor of \cref{sec:relsubdists} to show how it improves on rule~\eqref{eq:cobisim} of \cref{sec:cobisimilarity}.

\subsection{Example: Subdistributions}\label{sec:simplesubex}
We saw in \cref{sec:cobisimilarity} how the definition of cobisimilarity for coalgebras of the subdistribution functor does not allow us to easily produce proofs. Here, we show that our new notion of behavioural apartness is more usable in this regard.

For this, we first elaborate how the functor \(\D_s\) acts on the quotient map of an equivalence relation, i.e., a map \(q_R \colon X \twoheadrightarrow X/e(R)\) mapping an element to its equivalence class under the equivalence closure \(e(R)\) of \(R\).
The image of such a map under \(\D_s\) is given by
\begin{align*}
	\D_s(q_R)(\mu)([z]_R) & = \sum_{x \in {[z]}_R} \mu(x)
\end{align*}
where, by \({[z]}_R\) we mean the equivalence class of \(z\) under the equivalence closure \(e(R)\) of \(R\). We may also write this sum as \(\mu{[z]}_R\), an instance of the general notation \(\mu(S \subseteq X) = \sum_{s \in S} \mu(s)\).

Then rule~\eqref{eq:genintrule}, for some coalgebra \(\gamma \colon X \to \D_s(X)\), instantiates to
\begin{equation*}
	\AxiomC{\(\forall (x',y') \in R \ldotp x' \apart{} y'\)}
	\AxiomC{\(\exists z \in X \ldotp \gamma(x){[z]}_{\overline{R^s}} \neq \gamma(y){[z]}_{\overline{R^s}}\)}
	\BinaryInfC{\(x \apart{} y\)}
	\DisplayProof
\end{equation*}

Returning to the earlier concrete example of~\eqref{fig:subMC}, we can now produce a proof which closely matches the reasoning we suggested earlier. We start by noting that for states such as \(x_1\) and \(y_2\) with one state having no outgoing probability weight, we have the following proof of behavioural apartness:
\begin{align*}
	\AxiomC{\(R = \emptyset\)}
	\AxiomC{\(\gamma(x_1){[y_2]}_{\top_X} = 0 \neq 1 = \gamma(y_2){[y_2]}_{\top_X}\)}
	\BinaryInfC{\(x_1 \apart{} y_2\)}
	\DisplayProof
\end{align*}
where \(\top_X\) is the total relation on \(X\). In this same way, we can prove \(x' \apart{} y'\) for all pairs in the set \( S_1 = \{ (x_1, x_2), (x_1,y_2), (x_2,y_1), (y_1,y_2) \} \).
We go on to prove that both \(x\) and \(y\) are apart from both of \(x_1,y_1\). For example, we can prove
\begin{align*}
	\AxiomC{\(R = S_1\)}
	\AxiomC{\(\gamma(x){[x]}_{\top_X} = 1 \neq 0 = \gamma(y_1){[x]}_{\top_X}\)}
	\BinaryInfC{\(x \apart{} y_1\)}
	\DisplayProof
\end{align*}
Although we have already proved a number of apartness pairs, taking the apartness interior still yields the empty set, due to the missing pairs required to make the relation cotransitive. This means we can only distinguish states for which the total outgoing weight is different.

Only once we have proven apartness for all pairs in the set \(S := S_1 \cup \{ (x,x_1), (x,y_1), (y,x_1), (y,y_1) \}\) can we give the proof of \(x \apart{} y\):
\begin{align*}
	\AxiomC{\(R = S\)}
	\AxiomC{\(\gamma(x){[x_1]}_{\overline{R^s}} = 0.5 \neq 0.4 = \gamma(y){[x_1]}_{\overline{R^s}}\)}
	\BinaryInfC{\(x \apart{} y\)}
	\DisplayProof
\end{align*}
Note that here
\[
	\overline{R^s} = e(\overline{R^s}) = \{(x,y),(x_1,y_1),(x_2,y_2),(x,x_2),(y,y_2),(x,y_2),(y,x_2)\}^s \cup \Delta_X
\]
This demonstrates that we are indeed able to provide a proof of behavioural apartness using rule~\eqref{eq:genintrule} which is finite and built up from inequations exhibiting differences in transition weights. The situation can however still be improved, by reducing the work required in our proofs. As mentioned, we were required to prove for instance the apartness pair \(x \apart{} y_1\). As \(x\) does not contribute to the transition distributions of the states we are proving apart, this should not be necessary for the proof. Such proof obligations arise because we take the apartness interior/equivalence closure with respect to the entire state space, and (co)transitivity forces that we include many ``unnecessary'' pairs. In the next section, we show that we can restrict to apartness relations on just those states that are in the supports of the successor distributions, or more generally, states which are reachable in one step.

\section{Optimised Rules for Behavioural Apartness}\label{sec:optimisations}
To further improve our proof system, we show that it is sound and complete to prove apartness using an inductive step in which apartness need only be proved on states which are ``reachable in one step'' from the states which we are proving apart. We make this notion precise in the following definition.
\begin{definition}
	Given a coalgebra \(\gamma \colon X \to B(X)\), we will say that a subobject \(m \colon Z \subobj X\) is a one-step covering of a subobject \(s \colon S \subobj X\) if there exists a \(g \colon S \to BZ\) such that \(Bm \circ g = \gamma \circ s\).
\end{definition}
Using this definition, we show that inequality of pairs of states under the map \(Bq_{\overline{R}} \circ \gamma\) which we have used in all our rules so far, is equivalent to inequality under the map \(B q_{\overline{R}|_Z} \circ g\) where we restrict the relation \(R\) to a \(Z\) that is a one-step covering of some \(S\) via \(g\).
\begin{lemma}\label{lem:behapbase}
	Let \(\gamma \colon X \to B(X)\) be a \(\CSet\)-coalgebra, \(R \subseteq X \times X\) an equivalence relation, and \(S \subobj X\) a subobject. If there is a non-empty \(m \colon Z \subobj X\) which is a one-step covering of \(S\) via $g \colon S \rightarrow B(Z)$, then
	\[
		B q_R (\gamma (x)) \neq B q_R (\gamma (y)) \iff B q_{R|_Z} (g(x)) \neq B q_{R|_Z} (g(y))
	\]
	where $q_R \colon X \rightarrow X/R$, $q_{R|_Z} \colon Z \rightarrow Z/{R|_Z}$ are quotient maps on $X$, $Z$ respectively.
\end{lemma}
This means, in other words, that we can determine behavioural apartness by looking only at apartness on one-step coverings.
The given definition is a generalised version of the notion of the \emph{base of a functor} originally due to Blok~\cite{blokmsc} which can be stated as the smallest one-step covering. It has been shown that this notion indeed instantiates to one-step reachability~\cite{DBLP:conf/fossacs/BarloccoKR19,wißmann2020coalgebraic}.
As an example, for a coalgebra \(\gamma \colon X \to \D_s(X)\) for the subdistribution functor, the base for a singleton $\{x\} \subobj X$ is exactly the support of the successor distribution of \(x\).
More interestingly for our applications are one-step coverings of pairs of states \(x,y\), which will instantiate to the union of the supports of \(x\) and \(y\) in the subdistribution case.
In general, the base may not always exist. Indeed, it requires the base category to be complete and well-powered, and the behaviour functor to preserve wide intersections (see~\cite[Prop.~12]{DBLP:conf/fossacs/BarloccoKR19}). The restriction to finitary functors on \(\CSet\) (which we make in~\cref{thm:soundcomp}) however implies these conditions. Despite this, we choose not to take the base itself in the rule which we give in~\cref{thm:optrule}, as this makes the application of the rule and the proof of completeness simpler.

\begin{proof}
	Suppose we have a \(g\) such that \(\gamma \circ s = Bm \circ g\), with \(m \colon Z \subobj X\) non-empty.
	We form the restriction of a relation to the subobject \(m \colon Z \subobj X\) by the following pullback:
	\begin{equation*}
		\begin{tikzcd}
			R|_Z \arrow[r,tail] \arrow[d,tail] & Z \times Z \arrow[d,tail,"m \times m"] \\
			R \arrow[r,tail] & X \times X \arrow[phantom,from=1-1,to=2-2,very near start,"\lrcorner"]
		\end{tikzcd}
	\end{equation*}
	We then obtain the following diagram:
	\begin{equation*}
		\begin{tikzcd}
			R|_Z \arrow[r,shift left,"\pi_1^Z"] \arrow[r,shift right,swap,"\pi_2^Z"] \arrow[d] & Z \arrow[r,two heads,"q_R|_Z"] \arrow[d,tail,"m"] & Z/R|_Z \arrow[d,dashed,"!"] \\
			R \arrow[r,shift left,"\pi_1^X"] \arrow[r,shift right,swap,"\pi_2^X"] & X \arrow[r,two heads,"q_R"] & X/R
		\end{tikzcd}
	\end{equation*}
	where \(! \colon Z/R|_Z \to X/R\) is the unique map which arises from the fact that \(q_R \circ m\) coequalizes \(\pi_1^Z,\pi_2^Z\) by commutativity of the left square. We now consider the commutative diagram:
	\begin{equation*}
		\begin{tikzcd}[column sep=2ex]
			S \arrow[r,tail,"s"] \arrow[dr,swap,"g"] & X \arrow[r,"\gamma"] & B(X) \arrow[rr,"Bq_{R}"] & & B(X/R) \\
			& BZ \arrow[ur,swap,"Bm"] \arrow[rr,swap,"Bq_{(R)|_Z}"] & & B(Z/(R)|_Z) \arrow[ur,swap,"B!"] &
		\end{tikzcd}
	\end{equation*}
	This means that \(Bq_{R} \circ \gamma \circ s = Bq_{R} \circ Bm \circ g = B! \circ Bq_{(R)|_Z} \circ g\) so that
	\begin{align*}
		         & Bq_{R}(\gamma(s(x))) \neq Bq_{R}(\gamma(s(y)))                          \\
		\iff     & B! \circ Bq_{(R)|_Z} \circ g (x) \neq  B! \circ Bq_{(R)|_Z} \circ g (y) \\
		\implies & Bq_{(R)|_Z} \circ g (x) \neq Bq_{(R)|_Z} \circ g (y)
	\end{align*}
	Now, if \(B!\) is mono, the last implication becomes a bi-implication, which is what we want. In \(\CSet\) this holds as follows: we have \(!({[z]}_{R|_Z}) = {[z]}_R\) and it is clear that \({[z]}_R = {[z']}_R \implies {[z]}_{R|_Z} = {[z']}_{R|_Z}\), i.e., \(!\) is mono. Now, as \(Z\) is non-empty, \(Z/R|_Z\) is non-empty, so that \(!\) is split, and thus its monicity is preserved by \(B\).
\end{proof}

We are now in a position to give our final proof rule for behavioural apartness, so that we can prove behavioural apartness \(x \apart{} y\) with an inductive step, where only pairs of states in a one-step covering of \(\{x,y\}\) need to be proved apart.
\begin{theorem}\label{thm:optrule}
	Let \(\gamma \colon X \to B(X)\). For all \(x,y \in X\), and \(m \colon Z \subobj X\) a non-empty one-step covering of \(\{x,y\} \subobj X\) via $g \colon S \rightarrow B(Z)$, the following rule is sound and complete for behavioural apartness on \(\gamma\):
	\begin{align}\label{eq:apbase}
		\AxiomC{\(\forall (x',y') \in R \ldotp x' \apart{} y'\)}
		\AxiomC{\(g(x) \reca{R,Z}{B} g(y)\)}
		\BinaryInfC{\(x \apart{} y\)}
		\DisplayProof
	\end{align}
	where we define
	\[
		t_1 \reca{R,Z}{B} t_2 := B q_{e(\overline{R^s})|_Z} (t_1) \neq B q_{e(\overline{R^s})|_Z} (t_2)
	\]
\end{theorem}
Note that we now explicitly take the equivalence closure before restricting to \(Z\). This is required for the application of \cref{lem:behapbase} in the following proof.
\begin{proof}
	Suppose we have a proof tree with root \(x \apart{} y\) built using this rule involving a relation \(R\). Soundness holds by \cref{lem:behapbase} instantiated to the relation \(e(\overline{R^s})\), as for the same \(R\) this lemma tells us that \(\gamma(x) \reca{R}{B} \gamma(y)\) will hold, so that the premise of the sound rule~\eqref{eq:genintrule} holds.

	For completeness, we note that the entire state space \(X\) is always a non-empty one-step covering of \(\{x,y\}\) with \(g = \gamma\). The rule~\eqref{eq:apbase} then reduces to rule~\eqref{eq:genintrule}, which is complete.
\end{proof}

\begin{example}[Subdistributions]
We return to the example of coalgebras for the subdistribution functor, and show how rule~\eqref{eq:apbase} improves on rule~\eqref{eq:genintrule}.

First, note that we can specialise the rule to subdistributions:
\begin{align*}
	\AxiomC{\(\forall (x',y') \in R \ldotp x' \apart{} y'\)}
	\AxiomC{\(\exists z \in Z \ldotp g(x){[z]}_{E} \neq g(y){[z]}_{E}\)}
	\BinaryInfC{\(x \apart{} y\)}
	\DisplayProof
\end{align*}
where \(E = e(\overline{R^s})|_Z\).
For the example of~\eqref{fig:subMC}, we will again prove the apartness \(x \apart{} y\), using our new rule.
For the last proof step, we will use the one-step covering \(Z = \{x_1,x_2,y_1,y_2\}\) (the supports of \(x\) and \(y\)). Thus, we will have
\begin{align*}
	\AxiomC{\(R = \{(x_1,y_2), (x_1,x_2), (y_1,x_2), (y_1,y_2)\}\)}
	\noLine
	\UnaryInfC{\(g(x){[x_1]}_{E} = 0.5 \neq 0.4 = g(y){[x_1]}_{E}\)}
	\UnaryInfC{\(x \apart{} y\)}
	\DisplayProof
\end{align*}
where \(E = \{(x_1,y_1),(x_2,y_2)\} \cup \Delta_Z\).

We now still need to prove the apartness pairs in this \(R\). Here, we give one pair as an example; the rest are similar.
\begin{align*}
	\AxiomC{\(R = \emptyset\)}
	\AxiomC{\(g(x_1){[y_2]}_{E} = 0 \neq 1 = g(y_2){[y_2]}_{E}\)}
	\BinaryInfC{\(x_1 \apart{} y_2\)}
	\DisplayProof
\end{align*}
We have taken \(Z = \{y_2\}\) as one-step covering of \(\{x_1,y_2\}\). We see that while the proof step for such leaves does not change much when using the rule of \cref{thm:optrule}, the proof in its totality is easier to provide than what we showed in \cref{sec:simplesubex}, and fits better with the desired reasoning based on supplying witnesses of apartness and only reasoning about successors.
\end{example}

\begin{example}[Streams]
Taking \(B = A \times (-)\) for some set of symbols \(A\), we can instantiate the rules~\eqref{eq:genintrule} and~\eqref{eq:apbase} to stream systems, by first elaborating the premises. The condition in~\eqref{eq:genintrule} for a stream system \(\langle o,t \rangle \colon X \to A \times X\) becomes:
\begin{align*}
	     & \id_A \times q_{\overline{R^s}}(\langle o,t \rangle (x)) \neq \id_A \times q_{\overline{R^s}}(\langle o,t \rangle (y)) \\
	\iff & (o(x),q_{\overline{R^s}}(t(x))) \neq (o(y),q_{\overline{R^s}}(t(y)))                                                   \\
	\iff & o(x) \neq o(y) \lor \lnot(t(x) \mathrel{e(\overline{R^s})} t(y))                                                               \\
	\iff & o(x) \neq o(y) \lor t(x) \mathrel{(R^s)^\circ} t(y)
\end{align*}
In words, two states of a stream system behave differently if they have different outputs (the heads of the streams they represent are different) or their successors behave differently (the tails of the streams they represent are different).

Rule~\eqref{eq:genintrule} can now be instantiated to stream systems:
\begin{align*}
	\AxiomC{\(o(x) \neq o(y)\)}
	\UnaryInfC{\(x \apart{} y\)}
	\DisplayProof
	\qquad
	\AxiomC{\(\forall (x',y') \in R \ldotp x' \apart{} y'\)}
	\AxiomC{\(t(x) \mathrel{(R^s)^\circ} t(y)\)}
	\BinaryInfC{\(x \apart{} y\)}
	\DisplayProof
\end{align*}
In this form, the right-hand rule involving successors requires us to take the interior. As we have seen in \cref{sec:simplesubex}, this will be empty unless we have proven enough apartness pairs. Consider, for instance, the following simple example
\begin{equation}\label{eq:fibstream}
	\begin{tikzcd}
		x_0 \arrow[r,"1"] & x_1 \arrow[r,"1"] & x_2 \arrow[r,"2"] & x_3 \arrow[r,"3"] & x_4 \arrow[r,"5"] & \cdots
	\end{tikzcd}
\end{equation}
	in which \(o(x_i)\) is the \(i\)-th Fibonacci number, so that \(x_0\) generates the full stream of Fibonacci numbers and \(x_1\) generates the stream of Fibonacci numbers excluding the first one. It should be clear that these states are therefore behaviourally apart. However, in order to prove this with the above rules we must use the rule involving successors, and hence provide a useful relation \(R\). For this \(R\) to be an apartness on the whole state space and to be non-empty, we must already prove an infinity of apartness pairs (at least one including each of the states due to cotransitivity). Instantiating instead our optimised rule~\eqref{eq:apbase}, we obtain
\begin{align*}
	\AxiomC{\(o_g(x) \neq o_g(y)\)}
	\UnaryInfC{\(x \apart{} y\)}
	\DisplayProof
	\qquad
	\AxiomC{\(\forall (x',y') \in R \ldotp x' \apart{} y'\)}
	\AxiomC{\(t_g(x) \mathrel{\overline{E}} t_g(y)\)}
	\BinaryInfC{\(x \apart{} y\)}
	\DisplayProof
\end{align*}
where we again write \(E\) for \(e(\overline{R^s})|_Z\) and
where \(Z\) is some one-step covering of \(\{x,y\}\) via \(g\) and we write \(o_g\) and \(t_g\) for the components of \(g\).

	For the example~\eqref{eq:fibstream}, we can take the set \(Z = \{x_1, x_2\}\) which is a one-step covering of \(\{x_0,x_1\}\) via the map \(g \colon \{x_0,x_1\} \to A \times \{x_1, x_2\}\) defined by \(g(x_0) = (1,x_1)\) and \(g(x_1) = (1,x_2)\). The states \(x_1, x_2\) can be distinguished by their outputs, and so we have the following finite proof of apartness for \(x_0,x_1\), with \(\overline{E} = R^s\) so that \((t(x_0),t(x_1)) = (x_1,x_2) \notin E\):
\begin{align*}
	\AxiomC{\(R = \{ (x_1, x_2) \}\)}
	\AxiomC{\(o(x_1) \neq o(x_2)\)}
	\UnaryInfC{\(x_1 \apart{} x_2\)}
	\AxiomC{\(t(x_0) \mathrel{\overline{E}} t(x_1)\)}
	\TrinaryInfC{\(x_0 \apart{} x_1\)}
	\DisplayProof
\end{align*}
\end{example}

\subsection{Inductive Characterisation of \(\asymp\)}

In~\cite[Appendix A]{DBLP:journals/lmcs/GeuversJ21}, an inductive definition of cobisimilarity for Kripke polynomial functors is given. Here, we show the instantiation of the relation \(\reca{R}{B}\) to a class of functors extending the Kripke polynomial functors. This will allow us to easily obtain proof rules for coalgebras of functors in this class.

We consider a slight restriction of Kripke polynomial functors on \(\CSet\) as defined by Jacobs~\cite{DBLP:books/cu/J2016} extended with the subdistribution functor, the syntax of which we give using the following grammar:
\begin{align*}
	B ::= \Id \mid A \mid B \times B \mid B + B \mid B^A \mid \pow B \mid \D_s B
\end{align*}
where \(A\) is any set. The restriction is to binary coproducts, which matches the presentation in~\cite[Appendix A]{DBLP:journals/lmcs/GeuversJ21}. Note, also, that we do not restrict to \emph{finite} Kripke polynomial functors as we wish to cover the examples of LTSs and MDPs.

We can now instantiate the definition of the relation \(\reca{R}{B}\) given in~\eqref{eq:reca}, with the functor \(B\) built from the above grammar. We do this for \(\reca{R}{B}\) here to ease notation.
To replace \(\reca{R}{B}\) with \(\reca{R,Z}{B}\) we must take the apartness interior in the first statement with respect to \(Z\), and the complement in the case \(B = \D_s B_1\) must also be taken with respect to \(Z\). More precisely, we have that \(\overline{\reca{R,Z}{B_1}} = (Z \times Z) \setminus \reca{R,Z}{B_1}\).
The proof in each case is a routine calculation.
\begin{lemma}
	For a relation \(R \subseteq X \times X\) and functors $B, B_1, B_2 \colon \CSet \to \CSet$ we have the following inductive characterisation of $t_1 \reca{R}{B} t_2$.
	\begin{itemize}[itemsep=4pt]
		\item If \(B = \Id\), then \(t_1 \reca{R}{B} t_2 \iff t_1 \mathrel{(R^s)^\circ} t_2\).
		\item If \(B = A\), then \(t_1 \reca{R}{B} t_2 \iff t_1 \neq t_2\).
		\item If \(B = B_1 \times B_2\), then \((u_1,v_1) \reca{R}{B} (u_2,v_2) \iff u_1 \reca{R}{B_1} u_2 \lor v_1 \reca{R}{B_2} v_2\).
		\item If \(B = B_1 + B_2\), then \(t_1 \reca{R}{B} t_2 \iff [t_1, t_2 \in B_1 X \implies t_1 \reca{R}{B_1} t_2] \land [t_1, t_2 \in B_2 X \implies t_1 \reca{R}{B_2} t_2]\)
		\item If \(B = B_1^A\), then \(t_1 \reca{R}{B} t_2 \iff \exists a \in A \ldotp t_1(a) \reca{R}{B_1} t_2(a)\)
		\item If \(B = \pow B_1\), then \(t_1 \reca{R}{B} t_2 \iff [\exists u \in t_1 \ldotp \forall v \in t_2 \ldotp u \reca{R}{B_1} v] \lor [\exists v \in t_2 \ldotp \forall u \in t_1 \ldotp u \reca{R}{B_1} v]\)
		\item If \(B = \D_s B_1\), then \(t_1 \reca{R}{B} t_2 \iff \exists z \ldotp t_1 [z]_{\overline{\reca{R}{B_1}}} \neq t_1 [z]_{\overline{\reca{R}{B_1}}}\)
	\end{itemize}
\end{lemma}
As discussed by Sokolova in~\cite{DBLP:journals/tcs/Sokolova11}, the (sub)distribution functor can be seen as an instance of the finitely supported monoid valuations functor \(\mathcal{M}_{S,f}^-\), a generalisation of the finitely supported multiset functor (see also~\cite{DBLP:journals/tcs/Gumm09,DBLP:journals/entcs/GummS01}). Due to the construction of our proof system, rules for such a functor can be straightforwardly obtained by calculation of \(\reca{R}{\mathcal{M}_{S,f}^-}\).

\subsubsection{Example: LMPs}

There are of course now many examples for which we obtain a proof system. Here, we take coalgebras for the functor \(\D_s(-)^A\), with \(A\) some finite set of actions, sometimes called Markov Decision Processes (without rewards) or Labelled Markov Processes.

We consider the following LMP:
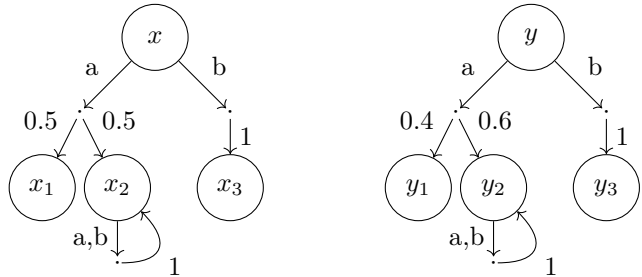
\begin{center}
	\begin{tikzpicture}[auto, on grid]
		\node[state] (x) {\(x\)};
		\node[inner sep=0pt] (xi1) [below left = of x] {\(\cdot\)};
		\node[inner sep=0pt] (xi2) [below right = of x] {\(\cdot\)};
		\node[state] (x1) [below left=1cm and 0.5cm of xi1] {\(x_1\)};
		\node[state] (x2) [below right=1cm and 0.5cm of xi1] {\(x_2\)};
		\node[inner sep=0pt] (x2i) [below=of x2] {\(\cdot\)};
		\node[state] (x3) [below=1cm of xi2] {\(x_3\)};
		\node[state] (y) [right=5cm of x] {\(y\)};
		\node[inner sep=0pt] (yi1) [below left = of y] {\(\cdot\)};
		\node[inner sep=0pt] (yi2) [below right = of y] {\(\cdot\)};
		\node[state] (y1) [below left=1cm and 0.5cm of yi1] {\(y_1\)};
		\node[state] (y2) [below right=1cm and 0.5cm of yi1] {\(y_2\)};
		\node[inner sep=0pt] (y2i) [below=of y2] {\(\cdot\)};
		\node[state] (y3) [below=1cm of yi2] {\(y_3\)};

		\path[->]
		(x) edge node[swap] {a} (xi1)
		edge node {b} (xi2)
		(xi1) edge node[swap] {\(0.5\)} (x1)
		edge node {\(0.5\)} (x2)
		(xi2) edge node {\(1\)} (x3)
		(x2) edge node[swap] {a,b} (x2i)
		(x2i) edge[out=0,in=-45,looseness=2] node[swap] {\(1\)} (x2.south east)
		(y2) edge node[swap] {a,b} (y2i)
		(y2i) edge[out=0,in=-45,looseness=2] node[swap] {\(1\)} (y2.south east)
		(y) edge node[swap] {a} (yi1)
		edge node {b} (yi2)
		(yi1) edge node[swap] {\(0.4\)} (y1)
		edge node {\(0.6\)} (y2)
		(yi2) edge node {\(1\)} (y3)
		;
	\end{tikzpicture}
\end{center}
States with no outgoing edge are those \(s\) for which \(\gamma(s)(\sigma) = 0\) (the zero distribution) for all \(\sigma \in A\).
Instantiating the generic rule~\eqref{eq:apbase} to this setting, gives
\begin{align*}
	\AxiomC{\(\forall (x',y') \in R \ldotp x' \apart{} y'\)}
	\AxiomC{\(\exists a \in A \ldotp \exists z \ldotp \gamma(x)(a)[z]_{e(\overline{R^s})|_Z} \neq \gamma(y)(a)[z]_{e(\overline{R^s})|_Z}\)}
	\BinaryInfC{\(x \mathrel{\apart{}} y\)}
	\DisplayProof
\end{align*}
We now take \(\sigma = \text{a}\), the one-step covering \(Z = \{x_1,x_2,x_3,y_1,y_2,y_3\}\), and
\[
	R = \{ (x_1,x_2), (x_2,x_3), (y_1,y_2), (y_2,y_3), (x_1,y_2), (x_2,y_3), (x_3,y_2), (x_2,y_3)\}
\]
Then, for example, \(z = x_2\) gives us
\begin{align*}
	\AxiomC{\(\vdots\)}
	\UnaryInfC{\(\forall (x',y') \in R \ldotp x' \apart{} y'\)}
	\AxiomC{\(\gamma(x)(a)[x_2]_{e(\overline{R^s})|_Z} = 0.5 \neq 0.6 = \gamma(y)(a)[x_2]_{e(\overline{R^s})|_Z}\)}
	\BinaryInfC{\(x \apart{} y\)}
	\DisplayProof
\end{align*}

\section{Future Work}\label{sec:future}
There are a number of avenues for future work. The first is based on the established link between notions of bisimilarity/behavioural equivalence (and their duals) and (coalgebraic) modal logic~\cite{DBLP:journals/tcs/Schroder08,DBLP:conf/icalp/FijalkowKP17,DBLP:conf/cmcs/0001MS20,DBLP:journals/lmcs/WissmannMS22,DBLP:conf/birthday/Geuvers22}. We would like to investigate how proofs of behavioural apartness in our system relate to formulas in a corresponding logic. Namely, can we extract formulas distinguishing states which are behaviourally apart? Such results already seem close by if we are able to choose our logic to match the reasoning present in our proof system. Indeed, a proof of behavioural apartness seems close to a proof that a formula is true in one state and its negation is true in the other.

A more recent notion of equivalence based on \emph{codensity lifting}~\cite{DBLP:journals/lmcs/KatsumataSU18,DBLP:journals/logcom/SprungerKDH21,DBLP:journals/ngc/KomoridaKHKHEH22} could also be an approach to more general notions of inequivalence. It has been shown how this can be used to define, e.g., behavioural preorders such as simulation~\cite{DBLP:journals/lmcs/KatsumataSU18} and also quantitative equivalences, which have further been linked to quantitative logics~\cite{DBLP:conf/lics/KomoridaKKRH21,DBLP:conf/concur/KonigM18,DBLP:conf/fossacs/GoncharovHNSW23,DBLP:conf/csl/Forster0HNSW23,DBLP:conf/csl/BeoharG0M23,DBLP:conf/stacs/BeoharG0MFSW24}. We would like to investigate whether the dual notions (which we may call codensity apartness) allow an easier development of expressive logics in these settings.
Part of this line of research should be to apply the ideas of \cref{sec:optimisations} to general liftings beyond that for behavioural apartness (also beyond \(\CSet\)). For instance, can we prove codensity apartness while only looking at states reachable in one step? We hope this would simplify the development of proof systems in the quantitative setting.
%
%
%
\bibliographystyle{splncs04}
\bibliography{bib}
%


\end{document}